# Mining Architecture Tactics and Quality Attributes Knowledge in Stack Overflow

**Tingting Bi** [a,c], **Peng Liang** [a,*], **Antony Tang** [b,d], **Xin Xia** [c]

[a] School of Computer Science, Wuhan University, 430072 Wuhan, China
[b] Faculty of Science, Engineering and Technology, Swinburne University of Technology, VIC 3122 Melbourne, Australia
[c] Faculty of Information Technology, Monash University, VIC 3166 Melbourne, Australia
[d] Software and Services Research Group, Vrije Universiteit Amsterdam, 1101, Amsterdam, The Netherlands
{bi_tingting,liangp}@whu.edu.cn, atang@swin.edu.au, xin.xia@monash.edu

**Abstract**
**Context**: Architecture Tactics (ATs) are architectural building blocks that provide general architectural solutions for addressing Quality Attributes (QAs) issues. Mining and analyzing QA-AT knowledge can help the software architecture community better understand architecture design. However, manually capturing and mining this knowledge is labor-intensive and difficult.
**Objective**: Using Stack Overflow (SO) as our source, our main goals are to effectively mine such knowledge; and to have some sense of how developers use ATs with respect to QA concerns from related discussions.
**Methods**: We applied a semi-automatic dictionary-based mining approach to extract the QA-AT posts in SO. With the mined QA-AT posts, we identified the relationships between ATs and QAs.
**Results**: Our approach allow us to mine QA-AT knowledge effectively with an F-measure of 0.865 and Performance of 82.2%. Using this mining approach, we are able to discover architectural synonyms of QAs and ATs used by designers, from which we discover how developers apply ATs to address quality requirements.
**Conclusions**: We make two contributions in this work: First, we demonstrated a semi-automatic approach to mine ATs and QAs from SO posts; Second, we identified little-known design relationships between QAs and ATs and grouped architectural design considerations to aid architects make architecture tactics design decisions.

## 1 Introduction

Software systems typically have multiple Quality Attributes (QAs) and design decisions are made to satisfy them. Architects make trade-off decisions to improve one QA to the detriment of another QA. Complex QA relationships, whilst known to experienced architects, are not well explored or documented. Apart from balancing inter-QA relationships, design decisions may sometimes involve the use of Architecture Tactics (ATs) [1]. AT aims to provide an established design to address a particular type of design problems with particular QA concerns. ATs serve as a building block of software architecture, and part of their purpose is to satisfy certain QAs. As opposed to architecture patterns which are related to multiple QAs, ATs are used for addressing one specific QA [3]. For example, ATs for performance, such as resource pooling, help to optimize response time (see an example from Stack Overflow[1] (SO) in Fig. 1). Furthermore, unlike design patterns that are described in terms of specific classes and associations, ATs are defined at a higher conceptual level of roles and responsibilities [2]. Tracing QAs and ATs can be useful for several reasons [32]. ATs can be analyzed in terms of QAs for understanding architectural design decisions, which can further enrich software and architecture documentation [36]. Documenting and understanding ATs and their rationale could be helpful for developers when they understand, implement, and modify the code of ATs for satisfying certain QAs [2][24].

[*] Corresponding author at: School of Computer Science, Wuhan University, China. Tel.: +86 27 68776137; fax: +86 27 68776027. E-mail address: liangp@whu.edu.cn (P. Liang).

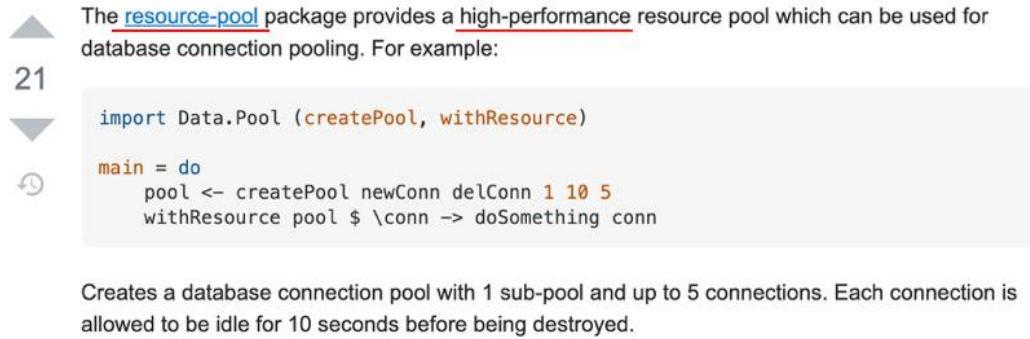

Fig. 1. An example of how ATs impact QAs from SO

Approaches of mining AT knowledge from specific software artifacts such as source code have been tried [2][6][7]. Some research focuses on understanding specific ATs and how the implementation of fault tolerance tactics affects architecture patterns [8]. Whilst these approaches can be valuable in helping developers to understand AT, little is known about the relationships between ATs and its impacts on QAs of a system. Whilst some of these AT-QA relationships are known in the industry, they are not commonly mentioned in research literature and books. We call them little-known QA-AT relationships and we intend to use machine learning algorithms to re-discover and highlight them. A better understanding of the characteristics of ATs and QAs as well as their inter-relationships would provide better and more tailored support for architects.

In addition, inexperienced architects sometimes find applying ATs to address QAs challenging mainly because of the numerous design decisions that need to be made in order to implement AT effectively [6]. In order to provide architects with such architectural knowledge, we need to build up this knowledge base by learning how ATs are used to address QA issues. To achieve this goal, we can gather and organise this knowledge from software discussion forums. In this work, we propose an approach for mining such architectural knowledge. We use Neural Language Model and machine learning techniques to train a dictionary-based classifier for the purpose of automatically mining the presence of ATs and QAs in online developer communities (i.e., Stack Overflow), and then we manually relate the ATs to relevant QAs to build a knowledge base of how developers use ATs. As such, our approach is designed to address architecture knowledge mining issues (e.g., ATs employed for addressing certain QAs) for undocumented AT decisions.

Our approach to knowledge mining is: firstly, we trained dictionary-based classifiers, which can be used for mining QA-AT posts from SO. Then we used the trained classifiers to mine more QA-AT posts from SO. We analyzed the mined posts for structuring an overview of QA-AT knowledge through understanding how developers apply ATs to address QAs in practice. Specifically, this study aims to address the following Research Questions (RQs):

---

[1] https://stackoverflow.com/

**RQ1:** Given our proposed semi-automatic knowledge mining approach, is it effective, in terms of accuracy (F-measure) and Performance (defined in Section 3.2.6), to mine QA-AT posts in SO?

**RQ2:** Applications of mined QA-AT knowledge.

**RQ2.1:** What are the common architectural design relationships between QAs and ATs that we can learn from the mined discussions?

**RQ2.2:** What design considerations can we provide to developers for making use of AT-QA relationships?

By answering the RQ1, we would be able to evaluate the effectiveness of our approach for mining QA-AT knowledge. The answers to RQ2 allow us to provide an overview of QA-AT knowledge through understanding how developers address QAs when using ATs. In particular, this work mainly has two contributions: (i) We proposed a semi-automatic approach, which can mine QA-AT posts in SO. Our approach can achieve an F-measure (0.865) by SVM with a trained dictionary to exploit term semantics for QA-AT posts mining. (ii) We conducted a qualitative analysis of the mined QA-AT posts for relating QAs and ATs. We also suggested a set of design considerations for developers to consider when using this QA-AT knowledge. Such knowledge can help developers make informed decisions of applying ATs to address certain QAs.

The remainder of the paper is organized as follow: Section 2 presents the motivation of this work. Section 3 gives the overview and details of each stage of our proposed approach. Sections 4 and 5 address the research questions and discuss the results, respectively. Section 6 describes the related works. Section 7 discusses the threats to validity. Finally, Section 8 concludes this work with future directions.

## 2 Motivation

Architects employ architectural frameworks, patterns, and tactics in design to address QA concerns such as performance, modifiability, maintainability etc. ATs are interrelated, it may be used with a complementary tactic or its use may exclude a conflicting tactic [14]. Since the application of AT, singly and in combination, influences the QA behaviours of a system, architects need to consider AT-QA knowledge appropriately [20].

Software development questions and answers (Q&A) sites (e.g., SO and R community[2]) gather knowledge that covers a wide range of topics [12]. These sites allow developers to share experience, offer help, and learn new techniques [16]. We provide two examples in Fig. 1 and Fig. 2, respectively, which show developers' concerns on implementing ATs in terms of certain QAs. SO is one of the most famous and popular online Q&A forums. It contains millions of posts contributed by tens of thousands of developers [7]. SO provides functions such as resurrecting and editing posts that can be inactive for long periods. It supports up voting competing answers and users can earn reputation points by posting interesting questions and answers [17]. Recent studies show that developers and architects use social media to discuss architecture-relevant information (e.g., features and domain concepts) [18][19]. In this work, we only considered the SO posts that have both questions and answers because through analyzing the questions and answers of the posts, we can explore what design problems developers had and what potential solutions they proposed. However, with a large volume of posts in SO, manually mining QA-AT knowledge is time-consuming and requires a lot of efforts. As such, applying semi-automatic approaches for mining QA-AT posts can significantly facilitate the tasks of finding the desired QA-AT knowledge, and doing that repeatedly. To this end, we decided to apply a semi-automatic approach to mine QA-AT knowledge in SO.

---

[2] https://community.rstudio.com/

Fig. 2. A QA-AT post from an issue tracking system

# 3 Knowledge mining approach

In this section, we describe our knowledge mining approach, including training data (i.e., relevant posts) collection and labelling, dictionary-based QA-AT post classifier training, and the process of empirical data analysis.

## 3.1 Overview of the knowledge mining approach

We proposed a knowledge mining approach, which comprises two stages: (a) Semi-automatic dictionary-based QA-AT post classifier training and (b) QA-AT posts mining and empirical analysis. An overview of our approach is provided in Fig. 3.

### *Stage 1: Semi-automatic dictionary-based QA-AT post classifier training*

ATs come in many different forms and can facilitate the betterment of QAs. For example, reliability tactics provide solutions for fault mitigation, detection, and recovery; performance tactics provide solutions for resource contention in order to optimize response time and throughput; and security tactics provide solutions for authorization, authentication, non-repudiation, and other such concerns. Finding a representative sample of ATs and how they impact QAs is far from trivial. In this stage, we trained QA-AT post classifiers, which can be used for mining QA-AT posts in SO. The execution process of *Stage 1* is composed of six steps that are described in Section 3.2.

### *Stage 2: QA-AT posts mining and empirical analysis*

ATs are measures taken to address software architecture quality attributes, or QAs, of a system. Using ATs, some QAs might improve whilst other QAs might be adversely affected. Bass and colleagues [3] discuss how the selection of tactics and design patterns relate to QAs. In our previous work [5], we analyzed the relationships between architecture patterns, QAs, and design contexts. In this work, we further explored the interactions between QAs and ATs which can help developers understand QA-AT relationships. The purpose of this stage is to mine more QA-AT discussions (i.e., posts) and investigate how developers discuss and apply ATs in terms of QAs. The execution of *Stage 2* is empirical analysis of the mined QA-AT posts that are described in Section 3.3.

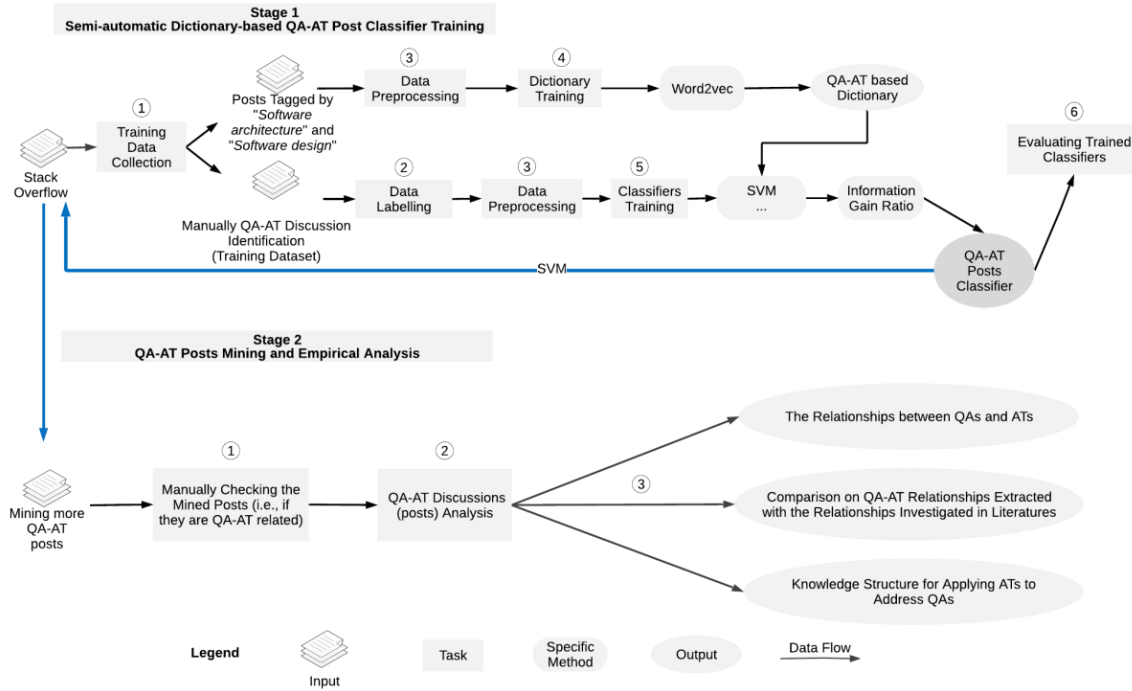

Fig. 3. An overview of the approach for QA-AT posts mining and analysis

## 3.2 Stage 1: Semi-automatic dictionary-based QA-AT post classifier training

### 3.2.1 Step 1: Data preparation

**Data preparation** is divided into two parts:

1. QA-AT posts collection for training classifiers: we applied the following criteria to select the QA-AT posts for training classifiers: a) posts need to be concerned with at least one of ATs; b) posts are related to at least one QA. We manually identified QA-AT posts and non QA-AT posts in SO, and our approach takes these posts as the training data for a QA-AT post classifier. We manually selected QA-AT posts using the tactic names of ***commonly*** used ATs and their relevant terms collected from [2][3][4][6][7] (see Table 1), and we list the collected ATs as below:

*Heartbeat*, *Audit trail*, *Resource pooling*, *Authentication*, *Scheduling*, *FIFO, Checkpoint*, *Rollback*, *Spare*, *Redundancy replication*, *Voting*, *Shadow operation*, *Secure session*, *Time out*, *Time stamp*, *Sanity checking*, *Functional redundancy*, *Analytical redundancy*, *Resisting attacks*, *Maintain data confidentiality*, *Recovering from attacks*.

About the QAs, we adopted the ISO 25010 standard that defines eight high-level QAs: Usability, Security, Reliability, Portability, Performance, Maintainability, Functional Suitability, and Compatibility [11]. We also referred to a wordlist[3], which is specified in the software engineering field for identifying QAs (see Table 2).

Note that, we (the first author and a master student) searched for relevant posts in their titles, tags, questions, comments, and answers of the posts that include QA and AT related terms. We retrieved 6,489 posts that contained relevant terms (see Tables 1 and 2). We then manually checked if the posts are QA-AT related, and finally we selected 1,165 QA-AT posts that include 1,203 QA-AT instances (see Table 3(a)).

---

[3] http://softwareprocess.es/y/neil-ernst-abram-hindle-whats-in-a-name-wordlists.tar.gz

Table 1. Selected architecture tactics with their related terms

| # | AT name | Related terms |
|---|---|---|
| AT1 | Heartbeat | heartbeat, ping, ping/echo, beat, decorator, piggybacking, outbound, period |
| AT2 | Audit trail | audit, trail, wizard, log, string, category, thread |
| AT3 | Resource pooling | pooling, pool, thread, connect, sparrow, processor, worker, time-wait, prototype, singleton, strategy, chain of responsibility, lazy load, static scheduling, dynamic priority scheduling |
| AT4 | Authentication | authentic, credential, challenge, login |
| AT5 | Checkpoint | checkpoint, checkpoints, barrier, weak point |
| AT6 | Rollback | layoff, restraint, austerity, abridgement, deliver |
| AT7 | Spare | spare, unoccupied, option, unused, logging, minutes |
| AT8 | Redundancy replication | redundancy replication, redundancy storage, zone-redundant, geo-redundant, replication |
| AT9 | Voting | voting, vote, balloting, choosing, voter, processor, preferred |
| AT10 | Shadow operation | shadow operation, shadow mode |
| AT11 | Secure session | secure session, security, removal |
| AT12 | Time out | time out, run out, constraint, action, monitor, timer, runtime |
| AT13 | Time stamp | time stamp, timestamp, time strap |
| AT14 | Sanity checking | sanity checking, sanity check |
| AT15 | Functional redundancy | functional redundancy, function requirement allocation |
| AT16 | Scheduling | schedule, dynamic priority scheduling, task, priority, adaptor, bridge, composite, flyweight, memento, observer, proxy, strategy |
| AT17 | FIFO | FIFO, first in first out |
| AT18 | Analytical redundancy | parallel, separate, warm restart, dual redundancy |
| AT19 | Resisting attacks | resisting attacks, detecting, detect, recovering, recover, sensor, authenticate, confidentiality, exposure, limit access, passwords, one-time, passwords, digital certificates |
| AT20 | Maintain data confidentiality | maintain data confidentiality, handle, protecting, routine, storage, mandatory |
| AT21 | Recovery from attacks | recovering from attacks, state, maintain, maintaining, redundant, access control, profile |

Table 2. Frequently discussed QAs and their related terms from mined QA-AT posts

| # | QA name | Related terms | Example |
|---|---|---|---|
| QA1 | **Performance (Efficiency)** | *performance, processing time, response time, resource consumption, throughput, efficiency, carrying into action, carrying out, operation, achievement, interaction, accomplishment, action* | "We propose the adaptive **heartbeat** between RM and NM to achieve a balance between updating NM's info promptly and minimizing the **response time** of extra heartbeats." |
| QA2 | **Maintainability** | *maintainability, update, modify, modular, decentralized, encapsulation, dependency, readability, interdependent,* | "How to adopt **pooling** to an existing object that has inline-field-initialization without sacrificing **code-maintainability and readability**." |
| | | *understandability, modifiability, modularity, maintain, analyzability, changeability, testability, encapsulation* | |
| QA3 | **Compatibility** | *compatibility, co-existence, interoperability, exchange, sharing* | "*I would like to be able to know about the **compatibility** of web service subscriptions to avoid duplicate request from distinct clients ... I needed built in browser ... and automatic **heartbeat** function offered by Stomp.js.*" |
| QA4 | **Usability** | *usability, flexibility, interface, user-friendly, default, configure, serviceability, convention, accessibility, gui, serviceableness, useableness, utility, useable, learnability, understandability, operability, function, use* | "*The aim of the **heartbeats** is to quickly find any nodes that go down, or if nodes can't communicate with the central server. **Usability** on the client nodes is an issue, so I don't want to use java (because that would require installing a jvm).*" |
| QA5 | **Reliability** | *reliability, failure, bug, resilience, crash, stability, dependable, dependability, irresponsibleness, recover, recoverability, tolerance, error, fails, redundancy, integrity, irresponsibleness, dependable, maturity, recoverability, accountability, answerableness* | "*I'm looking for a way in Python (2.7) to do HTTP requests with 3 requirements: **timeout** (for **reliability**) ... but none of them meet my requirements.*" |
| QA6 | **Functional Suitability** | *functional, function, accuracy, completeness, suitability, compliance, performing, employable, functionality, complexity, functioning* | "*Adding a formal interface for additional node **heartbeat** processing would allow admins to configure new **functionality** that is scheduler-independent without needing to replace the entire scheduler.*" |
| QA7 | **Security** | *security, safe, vulnerability, trustworthy, firewall, login, password, pin, auth, verification, protection, certificate, security system, law* | "*To ensure **security**, the **timeout** of the cookie is also set to 5 minutes, and my jquery performs a **heartbeat** back to the server to ensure the cookie doesn't expire.*" |
| QA8 | **Portability** | *portability, portable, cross platform, transfer, transformability, documentation, standardized, migration, specification, movability, moveableness, replaceability, adaptability* | "*Essentially I have a **portable** suite of windows 7 apps that are managed by a single backbone application. This backbone application handles monitoring the other apps for status and **heartbeat**.*" |

2. Posts collection for training a dictionary: we collected the posts tagged with "*software architecture*" or "*software design*"[4] to train the dictionary. One or multiple tags can be chosen by developers when they post a question in SO, and the tags indicate the topics of the posts. An example post tagged with "*software design*" used for training the dictionary is shown in Fig. 4. The output of dictionary training is ***a network*** of related words of QA and AT [30] together with the strength of the relationships between terms. For example, the terms "*throughput*" and "*scalability*" have a stronger semantic relationship than the terms "*throughput*" and "*agreement*". The trained dictionary extracts and makes use of the related terms for further improving QA-AT posts mining from SO. The process of dictionary training is detailed in Section 3.2.4. We excluded the posts that contain blocks of source code in the question part because most of such posts discuss programming problems [13]. Finally, we collected 2301 posts tagged with "*software architecture*" and "*software design*" to train the dictionary. Note that, these 2301 posts are different from the training data used in classifier training. Fig. 7 presents the experimental results with and without using the trained dictionary.

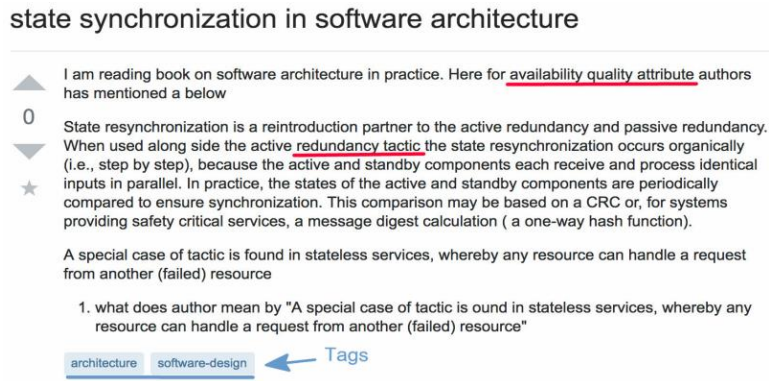

Fig. 4. An example of software architecture post for dictionary training.

In addition, to ensure the quality of the collected posts (i.e., two parts of training data collection), we only include the posts with at least one answer and positive scores.

### 3.2.2 Step 2: Data labelling

The manual labelling of QA-AT posts can be described as a multi-label binary classification process. A QA-AT post can be labelled under multiple labels if it is related to more than one QAs or ATs. Similar to the process of data collection, we first performed a pilot data labelling by three authors with 50 QA-AT posts in order to mitigate any personal bias in data labelling. In the formal data labelling, the QA-AT posts were manually labelled by two human annotators (i.e., the first author and one master student). After that, any disagreements on the labelled posts were discussed and confirmed with the second and third authors. To facilitate the manual labelling, we used MAXQDA[5], which is a tool for qualitative data analysis, to label the sentences of QA-AT posts. By the end of our labelling of the QA-AT posts, we made a final reliability test, and calculated Cohen's kappa reliability coefficient [37] for the categorization between the two annotators, which is 0.81. Note that this Cohen's kappa value was achieved after two rounds of data labelling within the formal data labelling, and the data labelling results have also been provided in our replication package [38].

After around three months of training data collection and labelling by the two annotators, we finally labelled 1,165 QA-AT posts for classifier training[6]. We retrieved AT posts by the keywords (see AT1 - AT21 in Table 3(a)), and each AT post returned is called retrieved AT instance (see the fourth column of Table 3(a)). We then checked if the retrieved AT instances discuss any QAs, and we included and labelled QA-AT instances (see the fifth column of Table 3(a)). This set of posts are used for classifier training and testing.

---

[4] The data for training dictionary can be found in [38] (i.e., data item 7).

[5] https://www.maxqda.com

[6] The data for training and testing classifiers can be found in [38] (i.e., data item 1).

A QA-AT post may discuss more than one ATs or QAs (e.g., participants discussed AT1 *Heartbeat* and AT13 *Time out* in one SO post). As such, a QA-AT post may contain one or more QA-AT instances. The number of labelled QA-AT instances found is 1,203 (see Table 3(b)) out of the 1165 posts.

For the training and testing dataset, we collected non QA-AT posts from SO manually. With two classes of posts (QA-AT and non QA-AT posts) in the dataset, the class imbalance problem has been known to hinder the learning performance of classification algorithms, and the standard machine learning algorithms yield better prediction performance with balanced datasets [46]. This work is an attempt to mine QA-AT posts with various machine learning algorithms, and consequently this is a balanced dataset in which the number of samples from the two classes are about the same (i.e., QA-AT and non QA-AT posts, see Table 3(b)). To enhance this dataset, 1,200 non QA-AT posts were collected by browsing the posts under the SO category "*software*" and labelled them as "non QA-AT" category Note that these 1,200 non QA-AT posts are additional data used for dictionary training in Step 4 (see Section 3.2.4). All the data and results of this study have been made available online [38].

Table 3(a). Information of labelled QA-AT instances for classifier training and testing (from 2012.01.01 to 2019.06.30)

|  | # | Architecture tactic | No. of retrieved AT instances | No. of labelled QA-AT instances |
|---|---|---|---|---|
| **QA-AT posts** | AT1 | Heartbeat | 521 | 128 |
|  | AT2 | Audit trail | 501 | 98 |
|  | AT3 | Resource pooling | 478 | 93 |
|  | AT4 | Authentication | 453 | 79 |
|  | AT5 | Checkpoint | 403 | 75 |
|  | AT6 | Rollback | 398 | 63 |
|  | AT7 | Spare | 387 | 61 |
|  | AT8 | Voting | 381 | 59 |
|  | AT9 | Redundancy replication | 354 | 57 |
|  | AT10 | Shadow operation | 289 | 54 |
|  | AT11 | Secure session | 281 | 50 |
|  | AT12 | Time out | 274 | 49 |
|  | AT13 | Time stamp | 270 | 47 |
|  | AT14 | Sanity checking | 261 | 46 |
|  | AT15 | Functional redundancy | 252 | 46 |
|  | AT16 | Scheduling | 221 | 27 |
|  | AT17 | FIFO | 200 | 38 |
|  | AT18 | Analytical redundancy | 197 | 42 |
|  | AT19 | Resisting attacks | 154 | 40 |
|  | AT20 | Maintain data confidentiality | 139 | 35 |
|  | AT21 | Recovering from attacks | 75 | 50 |
|  | Sum of labelled QA-AT instances | | | 1,200 |

Table 3(b). Information of labelled QA-AT posts and non QA-AT posts for classifier training and testing (from 2012.01.01 to 2019.06.30)

| # | Amount |
|---|---|
| No. of labelled QA-AT posts | 1,165 |
| No. of labelled non QA-AT posts | 1,200 |

With the 1,165 QA-AT posts as the training data, it is possible that some false positives are within the data (i.e., posts that contain some key terms regarding AT and QA but they are not actually QA-AT relevant). In order to check the validity of the data, we conducted another round of manual analysis of the 1,165 QA-AT posts to ensure that the data is correctly labelled.

### 3.2.3 Step 3: Data preprocessing

We take a number of steps to preprocess the posts: (1) Removing code snippets is to delete source code snippets that sometimes exist in the posts. (2) Tokenization is the process that breaks a stream of text up into words, phrases, symbols, or other meaningful elements called tokens. In our experiment, we only keep tokens that contain English letters. (3) Stop words removal: stop words are used often but carry little meaning to distinguish different categories of posts. We referred to a list of stop words, which contains a set of words (e.g., "the", "to", "of", "is"). Words that have a length of no more than two are also treated as stop words. (4) Stemming: the goal of stemming is to reduce inflected words to their word stem, base or root form.

### 3.2.4 Step 4: Dictionary training

In natural language processing, pre-trained word embeddings are used to alleviate the need for a large amount of task specific training data [43]. For example, QAs can be classified by applying word embeddings (i.e., terms matching) [39] on a set of keywords (i.e., related terms) to train a dictionary, and the trained dictionary can then be used to mine more QA-AT discussions. The process of dictionary training is shown in Fig. 5.

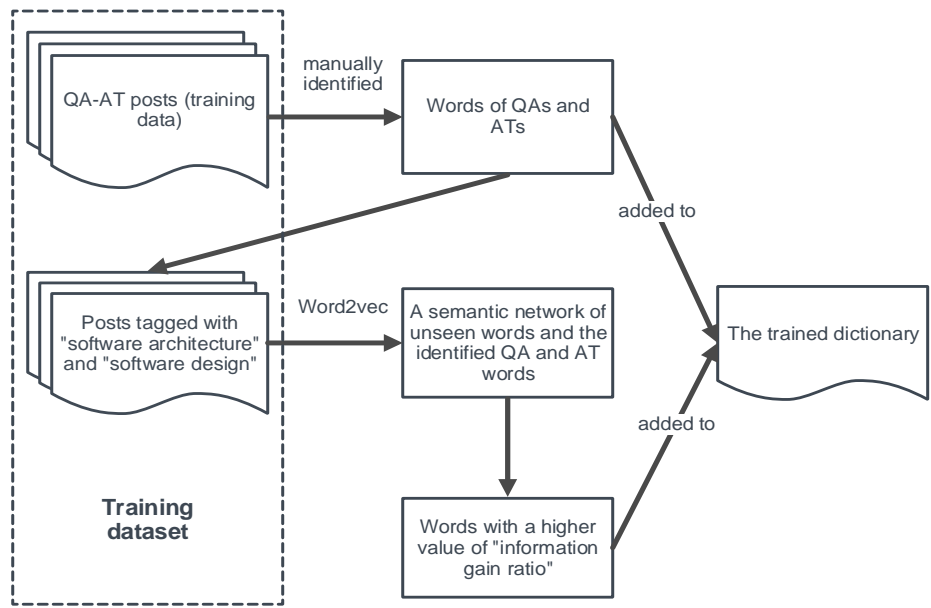

Fig. 5. The process of dictionary training.

Initially, some QA and AT terms were manually identified and added into the dictionary, and then some unseen related terms (also significantly contribute to QA-AT posts mining) were automatically extracted by Word2vec. We adopted an iterative process for extracting the keywords. In each iteration, two annotators went through each QA-AT post of training data for identifying related terms, and these terms were extracted and added to the dictionary. This process was repeated until no more related terms could be identified, and the manually identified QA and AT related terms are listed in Tables 1 and 2, respectively.

To cover unseen terms that can be used to mine more QA-AT posts, we used the posts tagged with "*software architecture*" or "*software design*" (collected in Step 1 as described in Section 3.2.1) to train a dictionary through constructing the semantic relationships between identified QA and AT words and unseen terms. We then applied the dictionary to train classifiers, which can mine more QA-AT posts from SO. In this work, we only used nouns to construct the semantic network of words, ignoring verbs, adjective, and adverbs. We employed the Word2vec tool, which provides a vector-based representation of words to get terms similarity by multiplying the vector of terms. A recent study shows that Word2vec provides a state-of-the-art performance for measuring words semantic similarity [27]. The semantic similarity between post $p_k$ and term $t_j$ is calculated based on the definition in [33], which is shown in Formula (1), in which $p_k$ denotes the QA-AT post $k$ expressed by a vector $p_k = (t_{k,1}, t_{k,i}, ..., t_{k,n})$, $t_{k,i}$ denotes term $i$ in $p_k$, $n$ denotes the number of terms in $p_k$, $w_{k,i}$ denotes the weight of term $t_{k,i}$, and $sim = (t_{k,i}, t_j)$ denotes the similarity between term $t_{k,i}$ and $t_j$, which is calculated by Word2vec. We included terms with values of $sim > 0.35$. For each post (for training the dictionary), we calculated all unique terms to get the similarity values between terms. The value of $i$ depends on the length of posts and is calculated by Formula (2). $\theta$ is a threshold increasing from 0 with an increment interval of 0.1. With the increase of $\theta$, the classification results (i.e., F-measure) have no obvious tendency, making it challenging to choose the value of $\theta$ which achieves the best classification result in F-measure [33]. Then we used Information Gain Ratio algorithm provided by the data mining tool Weka to re-sort the terms, which can be used for distinguishing QA-AT posts more effectively [31]. Gain Ratio measures the performance of a term to split the population of posts into two types of posts (i.e., QA-AT posts and non QA-AT posts). After comparing the values of Information Gain Ratio of words, we tried a set of values of Gain Ratio of words. To be specific, the values were selected from an intensity range from 0.100 to 0.800, and we empirically found that if the values of Information Gain Ratio of words are higher than 0.300, these words can achieve the best performance for QA-AT posts classification in terms of F-measure. Consequently, we added the unseen terms with an Information Gain Ratio value (> 0.300) into the dictionary [40].

$$sim(p_k, t_j) = \sum_{i=1}^{n}(w_{k,i} \times sim(t_{k,i}, t_j)) \tag{1}$$

$$N = \theta \times post\_length \tag{2}$$

### 3.2.5 Step 5: Classifier training

In this step, we used the manually labelled QA-AT posts to train the dictionary-based classifiers. We used a feature selection algorithms Word2vec and TF-IDF to select textual features and calculate the weight of features, and used these textual features to train a classifier [35]. We then used Information Gain Ratio to measure the ability of each word (i.e., the weight of features) of classifying the posts correctly into two types (i.e., this word is more unique or common for one particular type of posts). The range of Information Gain Ratio is between 0 and 1 and expresses the generative probability of each word with respect to the type of post (i.e., QA-AT and non QA-AT post) [34]. We applied six machine learning algorithms, Support Vector Machine (SVM), Bayes, Decision Tree (DT), Logistic Regression (LR), Random Forest (RF), and Bagging to train the classifiers.

To train the classifiers, 70% of the data (i.e., 1,165 QA-AT posts and 1,200 non QA-AT posts) is randomly selected as training set and the remaining 30% of the data as testing set (see Table 3(a)). The benefit of this technique is that it uses all the data for building the model, and the results often exhibit significantly less variance than those of simpler techniques such as holdout method. We used a library (i.e., scikit-learn) in Python V3.7. for training the classifiers, and we used default settings for each classifier[7] [47][48].

### 3.2.6 Step 6: Trained classifiers evaluation

We evaluated our approach that uses machine learning algorithms (i.e., SVM, Bayes, LR, DT, RF, and Bagging) with or without a trained dictionary on QA-AT posts mining. Precision is used to measure the exactness of prediction set, while recall evaluates the completeness. Precision and recall can be expressed mathematically, and in Formula (3) and (4), $TP$ denotes the number of posts classified as type QA-AT that are actually QA-AT; $FP$ denotes the number of posts classified as type QA-AT that are actually non QA-AT; $FN$ denotes the number of posts classified as type non QA-AT that are actually QA-AT; $TN$ denotes the number of posts classified as type non QA-AT that are actually non QA-AT. Please not that, as the training and testing sets are randomly selected, the results (i.e., precision, recall, and F-measure) of the classification by running the algorithms might be slightly different each time. We present the best results of our approaches in Section 4.2.1.

$$precision = \frac{TP}{TP+FP} \quad (3)$$

$$recall = \frac{TP}{TP+FN} \quad (4)$$

Based on precision and recall, we can calculate F-measure as below, which denotes the balance and discrepancy between precision and recall:

$$F - measure = 2 \times \frac{precision \times recall}{precision+recall} \quad (5)$$

As mentioned in Section 3.1, after getting the classifiers, we applied the trained classifiers to mine more QA-AT posts in SO, and we manually checked the mined posts whether they are really QA-AT related. We defined a metric to evaluated the classifiers (i.e., Performance):

$$Performance(\%) = \frac{correctly\_classified\_QA-AT\_posts}{total\_mined\_posts} \times 100\% \quad (6)$$

in which $total\_mined\_posts$ denotes the number of posts which are mined by the trained dictionary-based classifiers from SO, and $correctly\_classified\_QA-AT\_posts$ indicates the number of the true QA-AT posts, which were checked and confirmed by two researchers (i.e., the first author and a master student), and any disagreements on the QA-AT posts were discussed and resolved with the second author.

## 3.3 Stage 2: QA-AT posts mining and analysis

As shown in Fig. 3, in Stage 2, we trained and evaluated six dictionary-based classifiers (in *Stage 1*) to mine QA-AT posts in SO. Based on the most promising results through the highest F-measure out of the six algorithms (see Section 4.1), we selected and applied the dictionary-based classifier SVM.

To answer RQ2, we analyzed the mined QA-AT posts to identify the presence of QAs and ATs, and examined their relationships. We aimed to learn about developers' perception of QA-AT from their discussions. We employed constant comparison method [28] to analyze qualitative data (i.e., the mined QA-AT posts) for the purposes of: (1) identifying how developers discuss QAs and ATs (i.e., their presence, characteristic and nature); (2) comparing the relationships between the QAs and ATs that we have identified to the ones in the literature; and (3) identifying and classifying other topics (i.e., considerations) that are discussed by developers in the QA-AT posts.

---

[7] The machine learning source code can be found in [38] (i.e., experiments.py).

# 4  Results

## 4.1  Effectiveness of knowledge mining (Results of RQ1)

With RQ1, we investigated the effectiveness of our semi-automatic approach on QA-AT posts mining from two aspects: ***the results of trained dictionary*** and ***QA-AT posts mining from SO***.

**The results of trained dictionary**: To investigate the effectiveness of using the trained dictionary for improving QA-AT posts mining, we first evaluated the ability of the trained dictionary in accurately identification of QA and AT related words. The output of the dictionary is *a network* (i.e., semantic relationships) between the words manually identified by the authors and a set of other QA-AT related words extracted from the SO posts tagged with "*software architecture*" or "*software design*". Developers might use different words to describe QAs and ATs (i.e., not initially identified by the authors). In this work, we call those words "unseen terms". Including more relevant QA and AT words would be helpful to cover and mine more QA-AT posts in SO. In Section 3.2.4, we describe how we collect QA and AT related words for constructing the dictionary. The process of dictionary training starts with a set of seed words (i.e., QA and AT related words manually identified by the authors), and then unseen terms are added into the dictionary-based on the semantic relationships calculated by the values of similarity and Information Gain Ratio.

We provided an initial set of words which contain QA and AT related words, for example, *Time out* is an AT related word identified by the authors, and we calculated the values of similarities between *Time out* and unseen terms in the specific dataset (i.e., the collected posts tagged with "*software architecture*" or "*software design*"), and if the values of similarities between *Time out* and unseen terms are larger than 0.350, we include these unseen terms for further evaluation whether they should be added to the dictionary (e.g., the similarity value of "*Loadtime*" is 0.450). We then calculated the values of Information Gain Ratio of the unseen terms, if the values of Information Gain Ratio is larger than 0.300 (e.g., the Information Gain Ratio value of *Loadtime* is 0.427), the unseen terms (e.g., *Loadtime*) can be added to extend the dictionary, and a semantic relationship is created between, e.g., *Time out* and *Loadtime* (see Fig. 6). The dictionary training is an iteration process, and we then calculate the semantic relationship between the unseen terms which has been added into the dictionary and other unseen terms (e.g., *Loadtime* and *Modular* in Fig. 6) we calculate the semantic relationships between identified words (i.e., red nodes in Fig. 6).

An example of the output result of the dictionary is shown in Fig. 6, which is calculated and visualized by the Gephi tool. We used the red and grey circles to denote the words manually identified and the unseen terms extracted, respectively, and the calculated values of Similarities refer to the lines between the notes (i.e., semantic relationships between words) and the calculated values of Information Gain Ratio illustrate the notes in Fig. 6, we have not illustrated the complete dictionary in Fig. 6 due to the space limitation, and we made the completed results of the trained dictionary online [38].

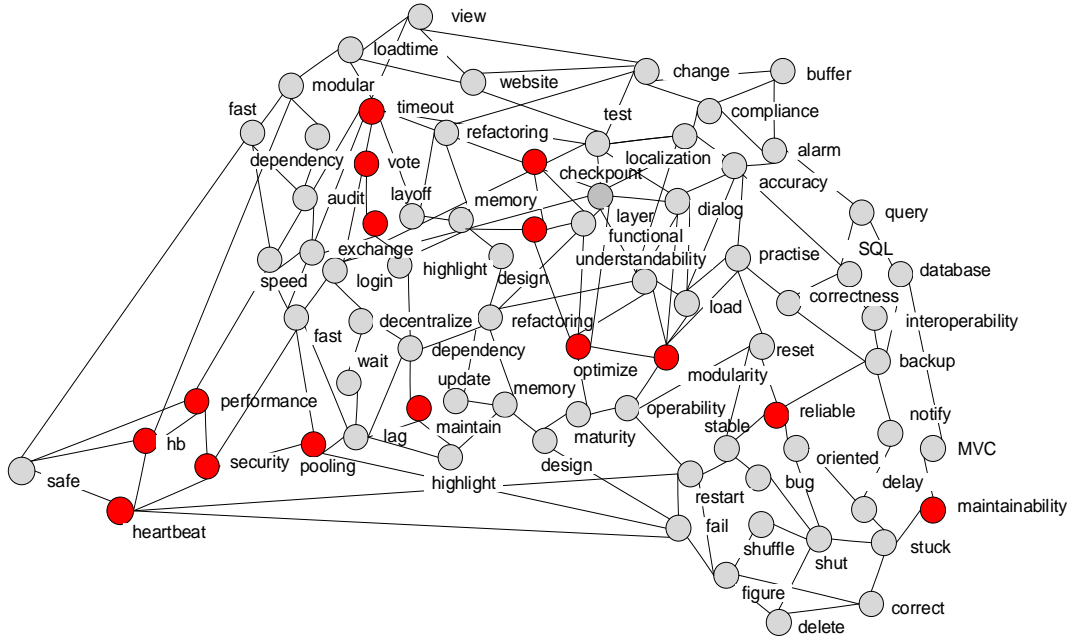

Fig. 6. Sematic relationships between terms of the trained dictionary

With the semantic relationships, we use Information Gain Ratio to calculate the values of unseen terms for QA-AT posts classification. We listed the top fifty unseen terms in Table 4. We observe that a set of unseen terms (not identified manually by the authors in the 1,165 QA-AT posts, i.e., not in Tables 1 and 2) are also related to QA and AT, and those unseen terms are helpful for improving QA-AT posts mining.

Table 4. Gain ratio of top fifty unseen terms of the dictionary[8]

| Gain ratio of unseen terms | | | | | | | |
|---|---|---|---|---|---|---|---|
| failure | 0.612 | throughput | 0.610 | monitor | 0.607 | evolution | 0.601 |
| penalty | 0.597 | scaling | 0.594 | congestion | 0.590 | selftest | 0.587 |
| buffer | 0.583 | response | 0.581 | component | 0.577 | protection | 0.571 |
| balancing | 0.569 | recovery | 0.564 | clone | 0.562 | update | 0.584 |
| integrity | 0.580 | replaceability | 0.579 | tolerate | 0.541 | restart | 0.512 |
| framework | 0.503 | prevention | 0.495 | sensor | 0.487 | transaction | 0.475 |
| operation | 0.471 | brokers | 0.469 | illegal | 0.467 | binding | 0.451 |
| model | 0.436 | prioritize | 0.429 | priori | 0.418 | loadtime | 0.427 |
| client | 0.423 | delay | 0.415 | tradeoff | 0.409 | interoperability | 0.403 |
| movability | 0.401 | optimize | 0.391 | useableness | 0.393 | collaborative | 0.391 |
| coupling | 0.386 | rest | 0.382 | microservices | 0.380 | mechanism | 0.375 |
| occur | 0.371 | timewait | 0.369 | modular | 0.365 | functionality | 0.361 |
| rollback | 0.360 | maptask | 0.358 | session | 0.351 | request | 0.348 |
| audit | 0.341 | wizard | 0.330 | simplify | 0.328 | query | 0.319 |
| wizard | 0.315 | periodic | 0.314 | loadbalancing | 0.312 | audit | 0.302 |

**QA-AT posts mining from SO**: We added the unseen terms (i.e., not identified by authors but related to QAs and ATs) into the training data for improving QA-AT posts mining. Fig. 7 shows a comparison of the experimental results with and without the trained dictionary on the testing dataset.

---

[8] The output of the training dictionary can be found in [38] (i.e., data item 4).

The results show that using the trained dictionary can consistently improve the six machine learning algorithms in terms of better weighted average F-measure for QA-AT posts mining. The improvements of average F-measures are: 19.9% with SVM, 21.7% with Bayes, 4.2% with Decision Tree (DT), 20.3% with Logistic Regression (LR), 8.8% with Random Forest (RF), and 12.8% with Bagging. In addition, the comparison of Recall, Precision, and F-measure values of the six machine learning algorithms for QA-AT posts mining is shown in Table 5. The highest F-measure (0.865) is achieved by SVM with the trained dictionary to exploit term semantics for QA-AT posts and non QA-AT posts mining.

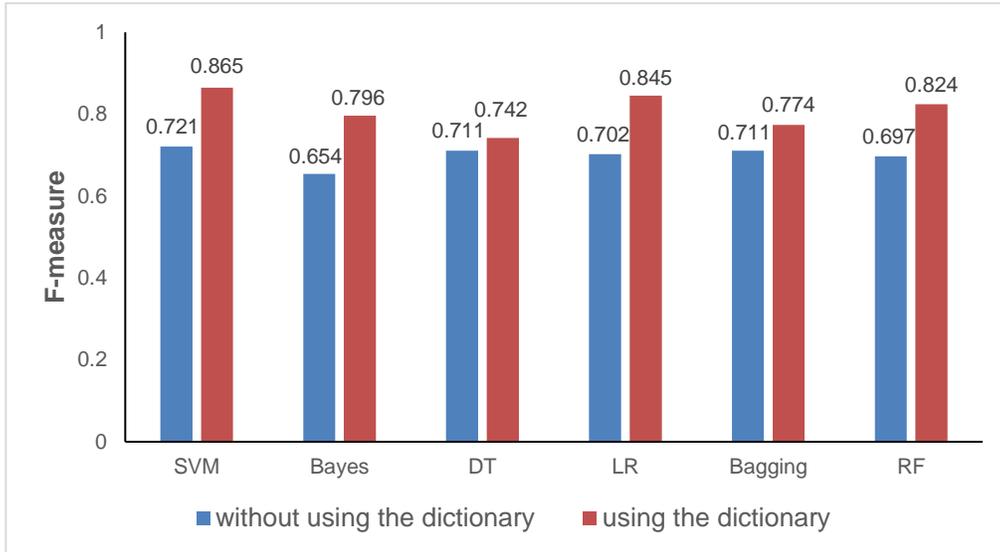

Fig. 7. Comparison of QA-AT posts mining results with and without using trained dictionary

Table 5. Results of QA-AT posts mining (with the trained dictionary)

|  | QA-AT post classification | | | | | | |
|---|---|---|---|---|---|---|---|
|  | True +ve (posts) | False +ve (posts) | True -ve (posts) | False -ve (posts) | Precision | Recall | F-measure |
| **SVM** | 903 | 20 | 1400 | 259 | 0.976 | 0.778 | 0.865 |
| **Bayes** | 831 | 163 | 1128 | 163 | 0.836 | 0.760 | 0.796 |
| **DT** | 829 | 242 | 959 | 335 | 0.774 | 0.712 | 0.742 |
| **LR** | 852 | 184 | 1016 | 313 | 0.822 | 0.731 | 0.845 |
| **Bagging** | 860 | 134 | 1178 | 193 | 0.865 | 0.816 | 0.774 |
| **RF** | 940 | 191 | 1025 | 209 | 0.831 | 0.818 | 0.824 |

As we described in Section 3.1, we applied the trained dictionary-based classifier (i.e., SVM algorithm) to mine more QA-AT posts in SO. We firstly limited the scope of crawled posts, and the process is similar to QA-AT posts collection (see Section 3.2.2). The crawled posts are tagged with at least one of AT terms (see Table 1). Note that the crawled posts are different from the training posts and we retrieved 12,761 crawled posts. Then we applied the trained dictionary-based classifier (using the SVM algorithm) to mine potential QA-AT posts from the set of crawled posts, and we found 5,103 posts. For the mined QA-AT posts from SO, two annotators (i.e., the first author and a master student) checked independently whether the mined posts are really QA-AT relevant, and any uncertain posts were discussed by three authors. Finally, 4,195 posts (out of the 5,103 mined posts) were manually checked and verified that are QA-AT relevant, and the value of **Performance** is 82.2%.

**RQ1 Summarization**: We used a set of metrics to evaluate the effectiveness of our approach: (1) The trained dictionary were used to identify the related terms and unseen terms of QA and AT in developers' discussions (see Tables 1 and 4). The **F-measure** values in Fig. 7 show that the trained dictionary can improve six algorithms on QA-AT posts mining. Some improvements are considerably more signficant (Bagging and RF) and some improvements are marginal (SVM, Bayes, and DT). (2) Our approach can reduce the manual efforts of mining QA-AT posts collection by human experts.

## 4.2 Applications of Mined Knowledge (Results of RQ2)

As mentioned in Section 2, the knowledge of ATs can provide solutions for addressing QA concerns. However, the relationships between ATs and QAs have not been explored systematically. To gather QA-AT knowledge and to help architects make informed design decisions when they apply ATs to address QAs in practice, we trained semi-automated dictionary-based classifiers (see Section 3.2), which can be used for mining QA-AT discussions from SO efficiently (see Section 4.1). The mined QA-AT knowledge was further empirically analyzed from two aspects to answer RQ2: (1) relationships between QAs and ATs (in Section 4.2.1), and (2) other key architectural design considerations discussed by developers when they apply ATs to address QAs (Section 4.2.2).

### 4.2.1 Results of RQ2.1: architectural design relationships between QAs and ATs

To systematically understand the QA-AT relationships, we present the results from the following three perspectives:

- **Interactions between various QAs and ATs**. We identified the presence of AT and QA instances and the interactions between various ATs and QAs in the mined 4195 QA-AT posts (see Fig. 8). We also identified the terms developers used to discuss QAs and ATs. We found that most of discussed QAs (i.e., about 45% QA-AT posts) describe QA behavioral properties of a system [41]. For example, a developer mentioned that "*Most unreleased resource issues result in general software reliability problems, but if an attacker can intentionally trigger a resource leak, it may be possible to launch a denial of service attack by depleting the resource pool[9]*", and in around 85% the mined QA-AT posts, developers discuss AT and QA issues using a variety of terms (see Tables 1 and 2), for example, developers used the words "*workload*", "*memory consumption*", "*application crash*", and "*low speed*" to describe Performance issues in the QA-AT posts. We counted the numbers of each QA and AT, and showed the interactions between various QAs and ATs in Fig. 8. The most frequently discussed QA and AT topics are **Performance** (1725 instances) and **Time out** (470 instances), respectively. In addition, the discussions on the interaction between Performance and several ATs (e.g., Time out and Checkpoint) are significantly higher than other QAs and ATs. We then investigated the architectural design relationships between various QAs and ATs. One or more ATs can be used to address the architectural design concerns of one or more QAs [26]. Such tactics have different levels of impacts on QAs. For example, developers mentioned that "*This scheduling is commonly adopted to improve system performance. For example, Scheduling services are used to execute jobs, including optimizing response time and latency*", and "*Fault detection tactic (heartbeat, Ping/Echo) is concerned with detecting a fault and notifying … (availability).*" Using the relationships that we have identified; architects and developers can select and calibrate the appropriate tactics to satisfy QAs.

- **Relationships between ATs and QAs in the mined QA-AT posts**. The objective of Stage 2 is to understand how ATs impact QAs in practice. We classified the influence as positive or negative (see Table 6, in which positive or "+" denotes that the AT benefits a specific QA, while negative or "-" shows that the AT hinders a specific QA [45]. If employing ATs is beneficial to certain QAs, we label the architectural design relationships between the ATs and QAs as "positive" (see the example of "positive" relationship between "Pooling" and "Performance" in Fig. 1). On the contrary, if applying ATs is a hindrance to certain QAs, the architectural design relationships between the ATs and QAs are labelled as "negative". Two annotators (i.e., the first author and a master student) read the mined QA-AT posts and labelled the relationships between QAs and ATs independently. Any controversial labels were further discussed with the second author. We tallied the numbers of relationships as "positive" or "negative". If developers did not make a point explicitly whether a specific QA is benefited or hindered by the ATs, we used "N/A" to denote the relationships. Please note that not all the interactions between ATs and QAs (see Fig. 8) are with an explicit negative or positive relationship. For example, developers do not explicitly discuss whether "Shadow operation" influences any QAs negatively or positively. Such QA-AT relationships are not shown in Table 6. The degree of positivity or negativity is the count of incidents we found in our samples.

- **Comparison on QA-AT relationships between the literature and SO**. To further investigate **RQ2.1**, we compared the QA-AT relationships in the mined QA-AT posts from SO (see Table 6) with the relationships from literature (i.e., the first author and a master student referred to software architecture books and literature) [3][4][6][10][14][15][21][26]. We explored which design relationships are documented in the literature and which design relationships are additional to the literature from the posts. We provide a comparison of the QAs with their related ATs from literature and additional design relationships that was mined from SO in Table 7.

---

[9] https://stackoverflow.com/questions/3673558/how-to-release-resource-after-delete-a-file-by-java

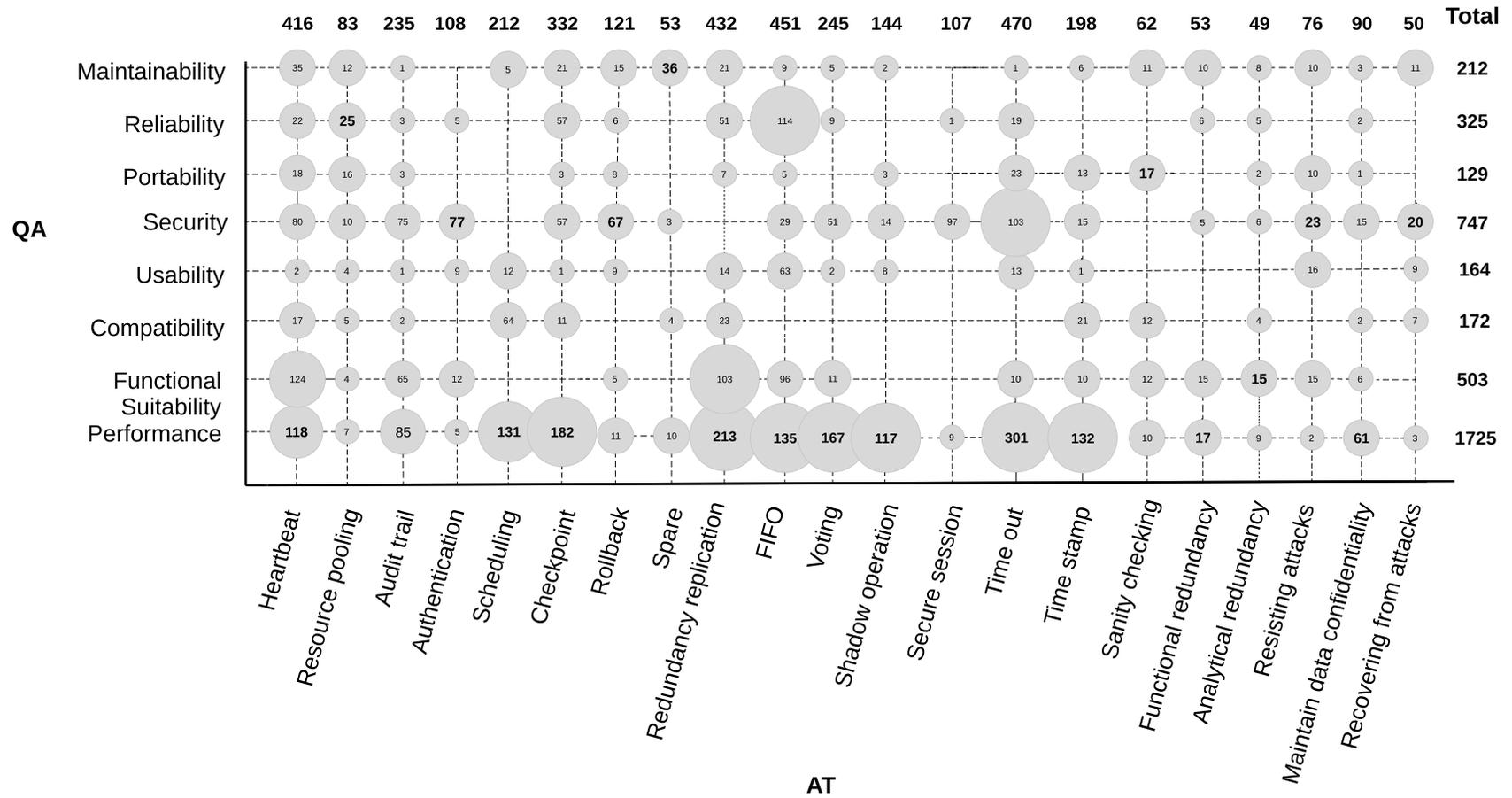

Fig. 8. Interactions between QAs and ATs in the mined QA-AT posts.

Table 6. Architectural design relationships between ATs and QAs in the mined QA-AT posts

|  | **Functional Suitability** | **Maintain-ability** | **Usability** | **Reliability** | **Performance** | **Compatibility** | **Security** | **Portability** |
|---|---|---|---|---|---|---|---|---|
| Time out | + (10) | N/A | + (5) | + (**17**) | + (**15**) | N/A | N/A | + (4) |
| Heartbeat | + (15) | + (1) | - (2) | + (10) | - (**47**) | + (1) | + (**28**) | + (**17**) |
| Time stamp | N/A | + (6) | N/A | N/A | - (2) | N/A | N/A | + (7) |
| Sanity checking | N/A | + (6) | N/A | N/A | - (1) | N/A | N/A | + (3) |
| Redundancy replication | + (7) | N/A | N/A | + (8) | + (12) | N/A | N/A | N/A |
| Functional redundancy | + (9) | + (9) | N/A | + (4) | + (12) | N/A | - (**3**) | N/A |
| Analytical redundancy | + (11) | N/A | N/A | N/A | + (4) | + (3) | + (4) | - (**1**) |
| Recovery from attacks | N/A | + (10) | - (**5**) | N/A | - (2) | + (**4**) | + (15) | N/A |
| Rollback | + (2) | + (**13**) | N/A | + (6) | + (5) | N/A | N/A | N/A |
| Scheduling | N/A | +(1) | N/A | N/A | + (**34**) | N/A | N/A | N/A |
| Checkpoint | N/A | N/A | N/A | N/A | + (6) | N/A | N/A | N/A |
| FIFO | N/A | N/A | + (**25**) | + (3) | + (10) | N/A | N/A | + (5) |
| Resource pooling | N/A | + (6) | N/A | + (5) | + (2) | N/A | N/A | + (1) |
| Secure session | N/A | N/A | N/A | N/A | N/A | N/A | + (4) | N/A |
| Resisting attacks | + (13) | - (**9**) | + (13) | N/A | N/A | N/A | + (1) | + (8) |
| Maintain data confidentiality | + (**21**) | N/A | N/A | + (5) | N/A | + (**4**) | N/A | N/A |
| Authentication | - (**6**) | N/A | + (3) | + (1) | N/A | N/A | + (14) | N/A |
| Voting | + (2) | N/A | + (1) | + (2) | + (7) | N/A | N/A | N/A |

Table 7. Comparison on the relationships between QAs and ATs documented in literature and additional relationships extracted from SO.

|  | Relationships between QAs and ATs from literature | Little-known relationships between QAs and ATs mined from SO |
|---|---|---|
| **Performance** | | |
| **Benefit to** *Performance* [3][6][10][14][21] | FIFO, Manage sampling rate, Limit event response, Reduce overhead, Bound execution times, Increase resource efficiency | Redundancy, Functional redundancy, Analytical redundancy, Rollback, Time out, Checkpoint, Resource pooling, Voting, scheduling |
| **Hinder to** *Performance* [3] | Heartbeat | Time stamp, Sanity check, Recovery from attacks |
| **Security** | | |
| **Benefit to** *Security* [3][4][10][14][21] | Detect service denial, Detect message delay, Authentication, Limit exposure, Heartbeat | Analytical redundancy, Resisting attacks, Recovery from attacks, Secure session |
| **Hinder to** *Security* | N/A | Functional redundancy |
| **Usability** | | |
| **Benefit to** *Usability* [3][10] | Maintain task model, Maintain user model, Maintain system model | Time out, FIFO, Resisting attacks, Authentication, Voting |
| **Hinder to** *Usability* [3][14] | Heartbeat | Recovery from the attacks |
| **Portability** | | |
| **Benefit to** *portability* [3] | Maintain task model, Maintain user model, Maintain system model | Time out, FIFO, Resisting attacks, Heartbeat, Time stamp, Sanity checking, Redundancy replication, Resource pooling, Recovery from attacks |
| **Hinder to** *portability* | N/A | Analytical redundancy |
| **Reliability** | | |
| **Benefit to** *Reliability* [3][10][15][21][26] | Heartbeat, Rollback, Voting, Exception, Redundancy Replication, Rollback | Time out, Functional redundancy, Resisting attacks, Recovery from attacks, Authentication, FIFO, Resource pooling, Maintain data confidentiality |
| **Functional suitability** | | |
| **Benefit to** *Functional suitability* | N/A | Time out, Heartbeat, Redundancy, Replication, Functional redundancy, Analytical redundancy, Rollback, Resisting attacks, Voting, Maintain data confidentiality |
| **Hinder to** *Functional suitability* | N/A | Authentication |
| **Maintainability** | | |
| **Benefit to** *Maintainability* | N/A | Heartbeat, Time stamp, Sanity checking, Functional redundancy, Rollback, Resource pooling, Recovery from attacks |
| **Hinder to** *Maintainability* | N/A | Resisting attacks |
| **Compatibility** | | |
| **Benefit to** *Compatibility* | N/A | Heartbeat, Analytical redundancy, Recovery from attacks, Maintain data confidentiality |

The architectural design relationships between ATs and QAs from mined posts are shown in Table 6. The comparison results between literature and SO are in Table 7, which reveal that: (1) Around 21% of the relationships between QAs and ATs extracted from SO are little-known relationships, for example, to our best knowledge, Time stamp can hinder Performance which has not been investigated in literature [3][4][6][10][14][15][21][24][26]. These little-known relationships can be added to literature to help developers consider potential impacts of using time-stamp when making trade-off decisions when they apply this AT; (2) An AT can affect multiples QAs simultaneously (see Table 7), for example, *Time out* can have an impact on five types of QAs (i.e., Functional Suitability, Performance, Usability, Portability, and Reliability). We further discuss these gaps between academia and industry on employing ATs to address QAs in Section 5.

### 4.2.2 Results of RQ2.2: architectural design considerations discussed in QA-AT posts

We analyzed the mined QA-AT posts to understand architectural design considerations between ATs and QAs. Whilst applying ATs to address QAs is well explored in existing works, e.g., [7][10], there is no guidelines for architects, who look for information on what considerations (e.g., design contexts) they need to consider when applying ATs to address QA concerns. As such, we analyzed the mined QA-AT posts using constant comparison method [28], which is a systematic approach to generate concepts and categories from the collected qualitative data, constantly compare incidents applicable to each category, and integrate categories and their properties, to explore and identify what design considerations developers discuss in the QA-AT posts. The first author coded and summarized a set of design considerations (i.e., architecture topics) from the mined QA-AT posts, and the results of coding were checked by the second author, finally any controversial results of coding and summarized topics of QA-AT posts were further discussed and resolved by the first three authors. For example, developers provided a brief background of the projects, what the design problems they had, and the design solutions they proposed in terms of the design problems.

The first author identified and coded the topics of the design discussions in the mined QA-AT posts, and after a discussion between the first two authors during selective coding, four main discussion topics were coded (i.e., Architecture pattern, Design context, Evaluation of design decision, and Tool support for monitoring QAs) in the collected QA-AT posts. We counted the percentages of QA-AT posts for each topic, for example, in around 47% posts (i.e., 1,975 out of the 4,195 QA-AT posts collected as the results of RQ1), developers discussed architecture patterns when they applied ATs to address QAs. An example of manual data coding using MAXQDA is shown in Fig. 9. The results of coding, examples of each topic, and percentages of related posts are listed in Table 8, and architects can use this architectural knowledge between QAs and ATs when designing.

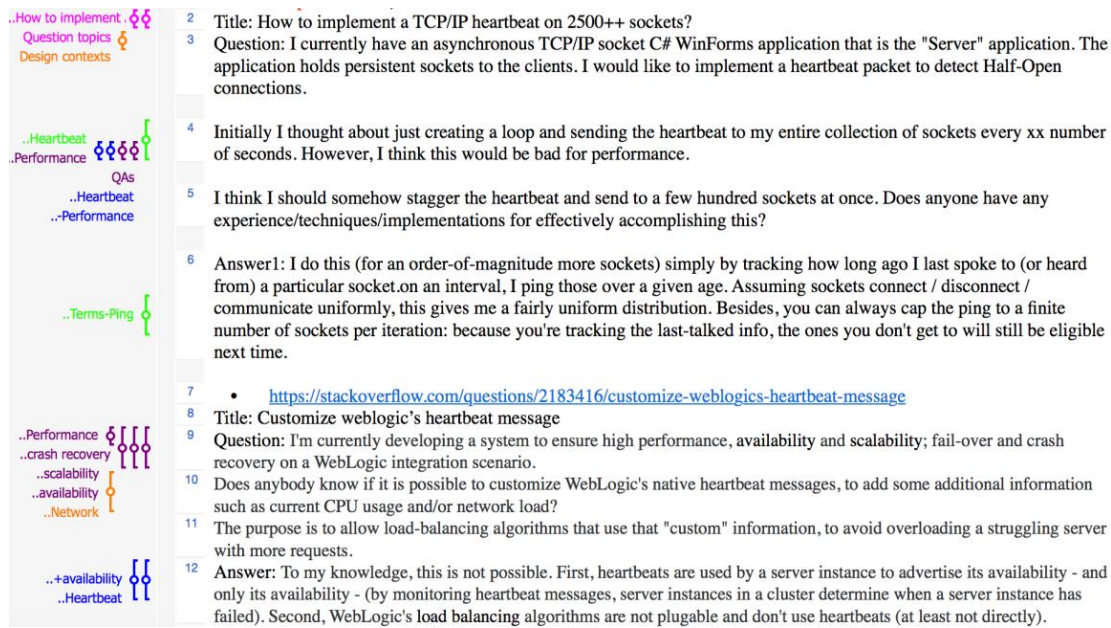

Fig. 9. An example of manual data coding using MAXQDA

Over half of the posts relate QAs to ATs (i.e., how ATs impact QAs). These discussions represent a set of design considerations of QA-AT. We group these considerations by discussion topics and sub-topics in Table 8. We first summarized the architecture design relationships between QAs and ATs in Table 6 (i.e., QA-AT architecture design relationships mined from SO) and Table 7 (i.e., little-known QA-AT architecture design relationships mined from SO compared with literature). Architects make trade-off decisions: whether to implement an AT that optimizes one QA to the detriment of another. Through these QA-AT related discussions, we explored trade-offs, design contexts and other issues that shape design decisions. The use of architecture patterns is one of the major discussion topics with 47% of QA-AT posts discussing this topic

About 28% of QA-AT posts discuss design contexts. Design contexts comprise the knowledge spanning the whole development lifecycle, which can be related to requirements, design decisions, and risks. In the posts, developers discuss design contexts when they make design decisions. The knowledge of the design context of specific scenarios influences the design decisions of applying ATs and architecture patterns to address certain QAs. The topics that they touched on include software, hardware, application and stakeholders. We have mined examples of them, as shown in Table 8. More examples can be found in [38].

Around 15% of QA-AT posts discuss design decision evaluation (e.g., developers compared alternatives of ATs to address specific QA concerns). Developers discuss reasons for achieving QAs and to predict system behaviour. Discussions on design alternatives also help select suitable ATs to achieve the desired QAs.

Finally, about 11% of QA-AT posts discuss how ATs can be applied in existing systems. In order to satisfy given QAs, architects want to apply and implement ATs in certain ways (e.g., Kafka as a message broker that implements *heartbeat* and *time out*). This design consideration is an important factor for developers when choosing and adopting the existing systems for applying ATs to address QAs in practice.

There are many and varied architectural design considerations that are useful to an architect. Our mining and research approach has allowed us to systematically identify and group some of these considerations by discussion topics. This identification process has allowed us to highlight architectural design patterns, design contexts, decision evaluation and AT applications are some of the main concerns of architects. Using this approach, knowledge can be continued to be mined and built-up to help architects use relevant architectural design knowledge.

Table 8. Architectural design considerations grouped by topics from the mined QA-AT posts

| Discussion topic | Subtopic | Example | Number of posts as a percentage to discussion topic |
|---|---|---|---|
| Architecture pattern | N/A | "There is the second approach of implementing a **Heartbeat** function to periodically check if the client responds. I do think this is the best approach for me / my scenario, but I am actually struggling with the implementation with **ASP.NET MVC**. How would I approach this in ASP.NET MVC?" | **47%** |
| Design context | Software context | "This is a classic problem with **Internet games and contests**. The simplest possible **attack against** your system is to run the HTTP traffic for the game through a proxy, catch the high-score save, and replay it with a higher score." | **28%** |
| | Hardware context | "The connection **pooling service** closes connections when they are not used; connections are closed every 3 minutes. The Decr Pool Size attribute of the ConnectionString property provides connection pooling service for the **maximum number of connections that can be closed every 3 minutes**." | |
| | Application domain | "For business information systems, **Security and Functionality are important**, and it's used by financial service companies for their **high performance requirements**." | |
| | Stakeholders | "All that said, an Access app with Jet/ACE back end can still perform well with more than **15/20 users** if those users are not in heavy data entry/editing mode. If there are mostly read-only user it's **pretty easy to support up to 50 users**." | |
| | Financial issues | "The **correct financing** is a process that requires the utmost attention to avoid the risks in software development." | |
| Evaluation of design decision | Improving certain QAs | "**Thread Pool management**: the ActorSystem is response for dispatching work from Actor instances to an underlying **thread pool**. If the ActorSystem has a more complete understanding of work distribution amongst your Actor set then it would be more efficient at allocating the thread pool's resources. However, the OS is pretty good resource allocation too so the **performance improvement should be negligible**." | **15%** |
| | Alternative | "I'm writing a method to check if there is new data in a **FIFO opened in RDONLY mode**. Until now I was using the poll() function but I realized that the kernel on which the code will run doesn't have this function and it implements a subset of the **functionality** and a subset of the POSIX **functionality**. There are **alternatives** to the poll function?" | |
| Application of ATs with existing systems | N/A | "I'm sure a few folks here have a similar use case of dealing with large **processing time** …Particularly, the recommended configuration setting around **heartbeat**, request **timeout**, max poll records, auto commit interval, poll interval, etc. if kafka is not the right tool for my use case, please let me know as well" | **11%** |

**RQ2 Summarization**: We extracted the relationships between QAs and ATs from SO and they are shown in Table 6. These relationships could help architects make decisions when they consider applying ATs to address QA concerns. Furthermore, we compared the extracted QA-AT relationships with the literature (Table 7) to analyze which QA-AT relationships were not reported in current literature. Through the comparison, we summarize the little-known QA-AT relationships (see Table 7) that can be used as a supplement to the literature.

In addition, the analysis performed in Section 4.2.2 shows that applying ATs to address QA concerns cannot be considered in isolation and the key considerations of architecture knowledge (see Table 8) would help architects to make design decisions when they apply ATs to address QAs.

Such considerations could help developers better understand two common design elements (i.e., AT and QA) and their interactions in practice. In addition, the popularity of the discussion topics and the considerations (in percentages) (see Table 8) suggest where attention can be placed.

# 5  Discussion

Although AT and QA are common architecture design elements [3][4][8], there is little knowledge on how ATs are used while trading off QAs in practice. QA-AT knowledge is typically unstructured and scattered in various resources (e.g., developer forums). Through mining and analyzing QA-AT knowledge from SO, a popular Q&A website for professional developers, we provide a guideline on the use of ATs with respect to QAs in practice. The main contributions of this work are: (1) Our approach (i.e., semi-automatic dictionary-based classifiers) can effectively mine QA-AT knowledge with an F-measure of 0.865 and the Performance is 82.2%, and 4,195 QA-AT posts (discussions) were mined from SO for empirical analysis; (2) Based on the empirical analysis of the mined QA-AT posts, we provided the relationships between QAs and ATs and a set of architectural design considerations that developers may consider when they address QAs using ATs in practice. The analyzed knowledge can help developers to understand the nature of QAs and ATs and apply ATs to address QAs. In this section, we further discuss and interpret the study results of each RQ.

**Semantic network of architectural knowledge (domain knowledge)**: The results of QA-AT post mining show that the trained dictionary is effective for making use of prior knowledge to construct semantic relationships between words and concepts (see Fig. 6 and Table 4). The trained dictionary results in better collection and representation of association on domain knowledge (i.e., architectural knowledge). We conclude that the semantic network of the words (i.e., domain knowledge) is effective for improving and facilitating QA-AT knowledge mining (see Fig. 7 and Table 5). However, as this work is an attempt for mining QA-AT knowledge, we only used 2, 365 architecture related posts (i.e., tagged with "*software architecture*" and "*software design*") to build the semantic network of architectural knowledge. We suggest that researchers and practitioners can employ more data on constructing the semantic network of architectural knowledge.

**Difficulties in AT and QA discussions extraction**: In this work, we mined QA-AT posts (i.e., discussions) for the purpose of understanding how developers apply ATs in terms of QAs. However, it is difficult to retrieve all QA-AT discussions by using words identification because developers may use different words (i.e. synonyms) from the trained dictionary to describe the same QA. Similarly, the words we used to extract AT discussions (see Table 1) may be insufficient for retrieving all AT discussions. Therefore, we need to employ multiple methods (e.g., deep learning techniques) for extracting more comprehensive QA-AT knowledge at different granularities (e.g., sentences and paragraphs) [49].

**The gap between academia and industry on employing ATs to address QAs**: ATs are fine grained reusable architectural building blocks and are widely used in practice. However, we found that there exists a gap between academia and industry applying ATs to address QAs. Very few researches introduce the negative impact of ATs on QAs. However, in SO, there are many cases in which certain types of ATs were mentioned with the characteristics of hindering specific QAs, such as Security could be hindered by Functional redundancy (see Table 7). Beyond that, we also found that there are few researches that investigate ATs for addressing certain QAs (i.e., Maintainability, Reusability, and Functional Suitability). We can only compare five QAs from ISO 25010 and their related ATs from literature and SO (see Table 7). In addition, around 21% little-known relationships between QAs and ATs are identified in SA. As such, this study can supplement what is currently lacking in the literature. For example, a set of ATs are extracted from SO that can be used to address Maintainability (see Table 7).

**Architectural design considerations in practice**: The analysis of the mined QA-AT posts (i.e., RQ2.2) have highlighted a number of architectural topics (see Table 8). Design considerations in QA-AT encompass the use of design patterns, design contexts, design decision evaluation and ATs in existing systems. Similar to QA-AT posts that have been mined, design discussions can reveal the trade-offs in decisions [42]. The result to RQ2.2 provides a glimpse on how developers deal with the interactions between QAs and ATs. Further research on the relationships between QAs and ATs in different design contexts can be useful.

# 6 Related work

There have been several attempts to provide methods and tools to assist architects to deal with QAs in architectural design. We report the literature in two areas: (a) using ATs to address QAs and (b) automatic architectural knowledge mining. We compare these works with our work in Table 9.

## 6.1 Applying architecture tactics to address quality attributes

Kim and colleagues proposed a quality-driven approach to address QAs using ATs. In their approach, ATs are represented as feature models, and their semantics are defined using Role-based Metamodeling Language (RBML) which is a UML-based pattern specification notation. Given a set of quality attribute, architecture tactics are selected and composed. There is a set of benefits of using this approach, for example, the variations captured in tactic specifications allow various tactic instantiations [14]. Bogner and colleagues investigated design decisions related to quality attributes for a Service-Based system. They proposed a lightweight manual design method called Service-Oriented Architecture Design Method (SOADM) that takes functional requirements and quality attributes as input and produces an architecture model of the necessary services and their interactions. To ensure that quality attributes goals are achieved, architectural tactics are used to enrich business services with system-related components that should realize the tactics [1]. Alashqar and colleagues introduced a new Multi Criteria Decision Making (MCDM) method for analyzing the preferences and interactions of quality attributes based on Choquet integral fuzzy measure. The analysis process is based on understanding the impact of implementing architecture tactics on quality attributes when developing an industrial system. These works are similar to our work that focuses on the relationships between ATs and QAs, while we mined and analyzed the knowledge of QA and AT from developers' discussions [25].

## 6.2 (Semi-) Automatic techniques in mining architectural knowledge

Mirakhorli and and colleagues evaluated and compared the efficacy of six classification algorithms (i.e., SVM, C45, Bagging, SLIPPER, Bayesian logistic regression, and AdaBoost) for identifying ATs from source code [2]. Mirakhorli and colleagues, in another piece of work, used classification techniques and information retrieval to identify architecture tactic-related classes in source code. This approach can be used to automatically construct traceability links between source code and architectural tactics. This approach also minimizes the human effort required to establish traceability that can be used to support maintenance activities and prevent architectural erosion [21]. Velasco-Elizondo and colleagues proposed an approach based on an information extraction technique (i.e., entities extraction) and knowledge representation (i.e., ontology) to automatically analyze architecture patterns considering specific quality attributes (e.g., Performance) [22]. To be specific, an ontology contains two sub-types of ontologies. One is English grammar-based ontology. The other is performance ontology that defines performance-specific concepts (e.g., security and throughput). Information extraction techniques (i.e., entity extraction) and the ontology were used to identify the relationships between architecture patterns and quality attributes in architecture pattern descriptions. The experiment results show that their approach is helpful for inexperienced architects to select architecture patterns through knowing whether specific quality attributes are promoted or inhibited. Casamayor and colleagues applied NLP techniques and K-means algorithm to semantically categorize candidate responsibilities into groups [23]. This approach firstly processes requirements documents by POS tagging technique to detect the actions and tasks that the system needs, then K-means is used to group similar responsibilities into architectural components. The experiments show that the results obtained by this approach correspond to the expected architectural components made by experts. These works motivate us to develop a semi-automatic approach to extract and mine QA-AT knowledge from textual information (i.e., SO).

## 6.3 Comparison between our work and related work

The works presented in Section 6.1 applied different approaches to mine AT knowledge and investigate the interactions between ATs and QAs, however, those works focus on specific ATs and they do not explore the relationships between ATs and QAs in practice. Furthermore, the work presented in Section 6.2 motived us to use a semi-automated approach to mine architectural knowledge at a larger scale and involving developers' opinions (i.e., from Stack Overflow).

We compare the characteristics of related work with our work in Table 9. Our work used SO for understanding how developers apply ATs to address QAs. We proposed a semi-automatic approach, which uses Neural Language Model for training the dictionary and machine learning techniques for training the QA-AT post classifiers. We then employed the trained QA-AT post classifiers to mine more QA-AT posts in SO, and further empirically analyzed the mined QA-AT posts for revealing their occurrences and the strengths of their relationships.

Table 9. Comparison of the characteristics of related work with our work

| Related Works | Data extraction approach | Data analysis approach | Focus | Artifacts Used |
|---|---|---|---|---|
| Our Work | A semi-automatic dictionary-based QA-AT posts extraction approach from SO | Descriptive statistics and constant comparison | Focus on relationships analysis between ATs and QAs. | 4,195 relevant posts form SO |
| Mirakhorli and colleagues [7] | Automatic source code extraction from OSS | Topic analysis | Focus on the relationships between topical domain concepts and the use of ATs. | Source code in 1,000 OSS projects |
| Mirakhorli and colleagues [6] | Automatic source code extraction from OSS | Semi-automatic data classification (i.e., machine learning) | Focus on how design patterns were used to implement various ATs. | Source code in 500 OSS projects |
| Mirakhorli and colleagues [2] | Automatic source code extraction from OSS | Manual analysis on classification results of machine learning and information retrieval (i.e., customized classifiers) | Focus on discovering and visualizing architectural code, and mapping these code segments to ATs. | Source code in 50 OSS projects |
| Harrison and colleagues [8] | Controlled experiment (i.e., two groups) | Analyzing and comparing experiment results from two groups manually | Focus on understanding the information of fault tolerance tactics that affect the architecture patterns of a system. | Information collected from two groups of participants |

| Gopalakrishnan and colleagues [9] | Automatic source code extraction from OSS | Topic analysis | Recommend ATs based on latent topics discovered in the source code. | Source code in 11,600 OSS projects |
| Sabry and colleagues [10] | Survey and questionnaire | Quantitative analysis | Focus on analyzing the relationships between QAs and ATs. | Data collected from a survey of 29 developers |
| Bi and colleagues [5] | Manual data extraction (i.e., relevant discussions) | Descriptive statistics and constant comparison | Focus on relationships extracting between architecture patterns, quality attributes, and design contexts. | 748 relevant posts (i.e., discussions) collected from SO |

# 7  Threats to validity

There are several threats that can potentially affect the validity of our research results. We discuss three threats to the validity according to the categorization in [29]. Internal validity is not considered since this study does not address any causal relationships between variables and results.

**Construct validity** focuses on whether the theoretical constructs are interpreted and measured correctly. A threat to construct validity in this study involves whether the training QA-AT posts used for experiments were labelled correctly by the researchers. To achieve a common understanding of various QAs and ATs, we reviewed literatures related to ATs and checked various terms that are synonyms with ATs. In addition, we used the definitions of QA types in the ISO 25010 standard. But using a standard cannot guarantee that the researchers understand the definitions of various QAs. To mitigate this threat, a pilot QA-AT posts extraction was conducted by three authors, and any disagreements on the extraction results were further discussed and resolved by the three authors, in order to get a consensus among researchers on the extraction of QA-AT posts. Another threat lies in the manual analysis of the mined QA-AT posts. To overcome this threat, we employed constant comparison method to analyze the mined QA-AT posts. The first author empirically analyzed the QA-AT posts, and the second author checked the results. Any disagreements on the coding results and analysis of QA-AT posts were discussed and resolved by three authors. Moreover, before the formal data analysis, we conducted a pilot data analysis by the first three authors, and any conflicting results were discussed and resolved to eliminate personal biases. Lastly, semi-automatic mining cannot retrieve all QA-AT posts. Our intention is to mine commonly used ATs and to understand the QA-AT knowledge discussed in SO. As such, missing ATs can be captured and added for training data collection in order to get more comprehensive results on QA-AT posts mining.

**External validity** refers to the extent of the generalizability of the study results. We only collected the data from SO. This may be a risk to the external validity of the results and findings, for example the extracted relationships between QAs and ATs (see Tables 6 and 7). However, since SO is the largest and most popular Q&A community widely used by software professionals worldwide [44], the risk of missing out representative data is mitigated. Moreover, QA-AT knowledge from other sources, like the development platform GitHub and social media Twitter are also needed critical to supplement our study results, which is considered as our future work to enhance external validity. Although we used constant comparison method to identify architecture design topics that architects are concerned with, the grouping of the data studied in RQ2.2 can be subjected to researchers' interpretations. Additionally, the data we used is limited to SO posts.

Whilst every measure is taken by the researchers to remain objective and thorough, our claim on the knowledge generalizability is still limited.

**Reliability** concerns with the repeatability of a study producing the same results. To mitigate the threats to reliability, we specified the process of our approach in a research protocol which can be repeated to produce similar results. The manual interpretations of the terms can be different for researchers with different architecture working experience. We mitigated this risk by working with these terms carefully. A pilot study was conducted by two authors and the analysis results were checked by three authors to eliminate the misinterpretation of the results.

# 8 Conclusions and future work

In this work, we proposed a semi-automatic approach to mine the knowledge of QAs and ATs from SO. This approach achieved an F-measure of 0.865 and Performance of 82.2% by using the dictionary-based machine learning techniques for mining the QA-AT posts in SO (see Section 4.2.1). Whilst the knowledge mining approach we employed is not new, its application to mine AT and QA knowledge is novel. In order to investigate how QAs are related to ATs and other architectural design considerations, we manually analyzed the mined QA-AT discussions. We used that data to see how ATs impact QAs in design.

We have several findings:

(1) we have developed and tested mechanisms to mine QA and AT knowledge effectively from unstructured architectural knowledge source SO. The mined data allow us to discover new architecture design terminologies. For example, developers used "*outbound*" or "*decorator*" to describe *Heartbeat*, which cannot be found from the literature. The synonyms or related concepts are shown in Tables 1 and 2;

(2) we have applied an empirical analysis method to relate QAs to ATs from the mined discussions. We have been able to see that different ATs have different impacts on QAs. Such relationships between QAs and ATs are new and useful. They could help architects consider quality requirements when selecting ATs;

(3) through the mining process and empirical analysis, we grouped the mined QA-AT posts by four architectural discussion topics (see Table 8) in which architects can consider when employing ATs.

With the findings, we conjecture that similar mining approaches can be further explored to extract software development knowledge from a variety of rich and unstructured developer discussion forums such as Stack Exchange, Bytes, and GitHub. It may be possible to use a similar mining approach to convert unstructured discussions into empirical- and evidence-based software engineering knowledge.

# Acknowledgements

This work is partially sponsored by the National Key R&D Program of China with Grant No. 2018YFB1402800. We would like to thank Tianlu Wang, who helped to collect and label the SO posts in this work.

# References

[1] J. Bogner, S. Wagner, and A. Zimmermann. Using architectural modifiability tactics to examine evolution qualities of service-and microservice-based systems, SICS Software-Intensive Cyber-Physical Systems, 34(2-3): 141-149, 2019.

[2] M. Mirakhorli and J. Cleland-Huang. Detecting, tracing, and monitoring architectural tactics in code, IEEE Transactions on Software Engineering, 42(3): 205-220, 2016.

[3] L. Bass, P. Clements, and R. Kazman. Software Architecture in Practice, 3rd Edition, Addison-Wesley Professional, 2012.

[4] N. B. Harrison and P. Avgeriou. How do architecture patterns and tactics interact? A model and annotation, Journal of Systems and Software, 83(10): 1735-1758, 2010.

[5] T. Bi, P. Liang, and A. Tang. Architecture patterns, quality attributes, and design contexts: how developers design with them? in: Proceedings of the 25th Asia-Pacific Software Engineering Conference (APSEC), Nara, Japan, pp. 49-58, 2018.

[6] M. Mirakhorli, P. Mäder, and J. Cleland-Huang. Variability points and design pattern usage in architectural tactics, in: Proceedings of the 20th International Symposium on the Foundations of Software Engineering (FSE), pp, 1-11, Article No. 52, 2012.

[7] M. Mirakhorli, J. Carvalho, J. Cleland-Huang, and P. Mäder. A domain-centric approach for recommending architectural tactics to satisfy quality concerns, in: Proceedings of the 3rd International Workshop on the Twin Peaks of Requirements and Architecture (TwinPeaks), Rio de Janeiro, Brazil, pp. 1-8, 2013.

[8] N. B. Harrison, P. Avgeriou, and U. Zdun. On the impact of fault tolerance tactics on architecture patterns, in: Proceedings of the 2nd International Workshop on Software Engineering for Resilient Systems (SERENE), London, United Kingdom, pp. 12-21, 2010.

[9] R. Gopalakrishnan, P. Sharma, M. Mirakhorli, and M. Galster. Can latent topics in source code predict missing architectural tactics? in: Proceedings of the 39th International Conference on Software Engineering (ICSE), Buenos Aires, Argentina, pp. 15-26, 2017.

[10] A. E. Sabry. Decision model for software architectural tactics selection based on quality attributes requirements, Procedia Computer Science, 65: 422-431, 2015.

[11] ISO, ISO/IEC 25010, Systems and software engineering – Systems and software Quality Requirements and Evaluation (SQuaRE) – System and software quality models, pp. 1-34, 2011.

[12] R. Abdalkareem, E. Shihab, and J. Rilling. What do developers use the crowd for? a study using stack overflow, IEEE Software, 34(2): 53-60, 2017

[13] S, Nasehi, J. Sillito, F. Maurer, and C. Burns. What makes a good code example?: A study of programming Q & A in stack overflow, in: Proceedings of the 28th IEEE International Conference on Software Maintenance (ICSM) Trento, Italy, pp. 25-34, 2012.

[14] S. Kim, D. Kim, L. Lu, and S. Park. Quality-driven architecture development using architecture tactics, Journal of Systems and Software, 82(8): 1211-1231, 2009.

[15] N. B. Harrison and P. Avgeriou. Incorporating fault tolerance tactics in software architecture patterns, in: Proceedings of the 2008 RISE/EFTS Joint International Workshop on Software Engineering for Resilient Systems (SERENE), Newcastle Upon Tyne, UK, pp. 9-18, 2008.

[16] J. Vassileva. Toward social learning environments, IEEE Transactions on Learning Technologies, 1(4): 199-214, 2008.

[17] B. Vasilescu, A. Capiluppi, and A. Serebrenik. Gender, representation and online participation: A quantitative study of stackoverflow, Interacting with Computers, 6(5): 488-511, 2013.

[18] M. Soliman, M. Galster, A. R. Salama, and M. Riebisch. Architectural knowledge for technology decisions in developer communities: An exploratory study with stackoverflow, in: Proceedings of the 13th Working IEEE/IFIP Conference on Software Architecture (WICSA), Venice, Italy, pp. 128-133, 2016.

[19] D. Pagano and W. Maalej. How do open source communities blog?, Empirical Software Engineering, 18(6): 1090-1124, 2013.

[20] T. Bi, P. Liang, A. Tang, and C. Yang. A systematic mapping study on text analysis techniques in software architecture, Journal of Systems and Software, 144: 533-558, 2018.

[21] M. Mirakhorli, Y. Shin, J. Cleland-Huang, and M. Cinar. A tactic-centric approach for automating traceability of quality concerns, in: Proceedings of the 34th International Conference on Software Engineering (ICSE), Zurich, Switzerland, pp. 639-649, 2012.

[22] P. Velasco-Elizondo, R. Marín-Piña, S. Vazquez-Reyes, A. Mora-Soto, and J. Mejia. Knowledge representation and information extraction for analyzing architectural patterns, Science of Computer Programming, 121: 176-189, 2016.

[23] A. Casamayor, D. Godoy, and M. Campo. Functional grouping of natural language requirements for assistance in architectural software design, Knowledge-Based Systems, 30(6): 78-86, 2012.

[24] F. Bachmann, L. Bass, and R. Nord. Modifiability Tactics, Software Engineering Institute, Technical report, Carnegie Mellon University, Pittsburgh, 2007.

[25] A. M. Alashqar, A. A. Elfetouh, and H. M. El-Bakry. Analyzing preferences and interactions of software quality attributes using choquet integral approach, in: Proceedings of the 10th International Conference on Informatics and Systems (ICIS), Giza, Egypt, pp. 298-303, 2016.

[26] N. B. Harrison and P. Avgeriou. Implementing reliability: the interaction of requirements, tactics and architecture patterns, Architecting dependable systems VII. Springer, Berlin, Heidelberg, pp. 97-122, 2010.

[27] T. Mikolov, K. Chen, G. Gorrado, and J. Dean. Efficient estimation of word representations in vector space, in: Proceedings of the 1st International Conference on Learning Representations (ICIL), Scottsdale, Arizona, USA, pp. 1128-1135, 2013.

[28] B. G. Glaser and A. L. Strauss. The Discovery of Grounded Theory: Strategies for Qualitative Research, Transaction Publishers, 2009.

[29] M. Höst, P. Runeson, M. C. Ohlsson, B. Regnell, and A. Wesslén, Experimentation in Software Engineering. Springer, 2012.

[30] Y. Li, Z. A. Bandar, and D. McLean. An approach for measuring semantic similarity between words using multiple information sources, IEEE Transactions on Knowledge and Data Engineering, 15(4): 871-882, 2003.

[31] J. Dai and Q. Xu. Attribute selection based on information gain ratio in fuzzy rough set theory with application to tumor classification, Applied Soft Computing, 13(1): 211-221, 2013.

[32] D. Falessi, G. Cantone, and R. Kazman. Decision-making techniques for software architecture design: A comparative survey, ACM Computing Surveys, 43(4): 1-28, 2011.

[33] C. H. Li, J. C. Yang, and S. C. Park. Text categorization algorithms using semantic approaches corpus-based thesaurus and WordNet, Expert Systems with Applications, 39(1): 765-772, 2012.

[34] J. R. Quinlan. C4.5: Programs for Machine Learning, San Francisco, CA, USA: Morgan Kaufmann Publishers Inc., 1993.

[35] G. Forman. An extensive empirical study of feature selection metrics for text classification, Journal of Machine Learning Research, 3(3): 1289-1305, 2003.

[36] W. Ding, P. Liang, A. Tang, and H. van Vliet. Knowledge-based approaches in software documentation: A systematic literature review, Information and Software Technology, 56(6): 545-567, 2014.

[37] J. Cohen. A Coefficient of Agreement for Nominal Scales, Educational and Psychological Measurement, 20(1): 37, 1960.

[38] Mining Architecture Tactics and Quality Attributes Knowledge in Stack Overflow: Replication Package: https://github.com/QA-AT/Mining-QA-AT-Knowledge-in-SO

[39] J. Cleland-Huang, R. Settimi, X. Zou, and P. Solc. The detection and classification of non-functional requirements with application to early aspects, in: Processings of the 14th IEEE International Requirements Engineering Conference (RE), pp. 36-45, 2006.

[40] A. G. Karegowda, A. S. Manjunath, and M. A. Jayaram. Comparative study of attribute selection using gain ratio and correlation based feature selection, International Journal of Information Technology and Knowledge Management, 2(2): 271-277, 2010.

[41] J. Eckhardt, A. Vogelsang, and D. M. Fernández. Are "non-functional" requirements really non-functional? an investigation of non-functional requirements in practice, in: Proceedings of the 38th International Conference on Software Engineering (ICSE), pp. 14-22, Austin, TX, USA, 2016.

[42] A. Alebrahimand and M. Heisel. Bridging the Gap Between Requirements Engineering and Software Architecture: A Problem-Orieneted and Quality-Driven Method. Springer, 2017.

[43] S. Godbole, I. Bhattacharya, and A. Gupta. Building re-usable dictionary repositories for real-world text mining, in: Proceedings of the 19th ACM international conference on Information and knowledge management (CIKM), pp. 1189-1198, Toronto, Ontario, Canada, 2010.

[44] S. Meldrum, S. A. Licorish, and B. T. R. Savarimuthu. Crowdsourced knowledge on stack overflow: a systematic mapping study, in: Proceedings of the 21st International Conference on Evaluation and Assessment in Software Engineering (EASE), pp. 180-185, Karlskrona, Sweden, 2017.

[45] N. B. Harrison and P. Avgeriou. Leveraging architecture patterns to satisfy quality attributes, in: Proceedings of the 1st European Conference on Software Architecture (ECSA), pp. 263-270, Aranjuez, Spain, 2007.

[46] S. Kotsiantis, D. Kanellopoulos, and P. Pintelas. Handling imbalanced datasets: A review, GESTS International Transactions on Computer Science and Engineering, 30(1): 25-36, 2006.

[47] G. A. A. Prana, C. Treude, F Thung, T. Atapattu, and D. Lo. Categorizing the content of GitHub README files, Empirical Software Engineering, 24(3): 1296-1327, 2019.

[48] C. Treude and M. P. Robillard. Augementing API documentation with insights from Stack Overflow, in: Proceeding of the 38th IEEE International Conference on Software Engineering (ICSE), pp. 392-403, Austin, TX, USA, 2016.

[49] R. Witte and Q. Li. Text mining and software engineering: an integrated source code and document analysis approach, IET Software, 2(1): 3-16, 2008.